\documentclass[%
preprint,
showpacs,
 amsmath,amssymb,
 aps,
pre,
]{revtex4-1}

\usepackage{graphicx}

\usepackage{dcolumn}
\usepackage{bm}

\begin{document}

\preprint{}

\title{Renormalization and universality of blowup \\ in hydrodynamic flows}

\author{Alexei A. Mailybaev}
\affiliation{%
Instituto Nacional de Matem\'atica Pura e Aplicada -- IMPA, Rio de Janeiro,
Brazil
\footnote{Estrada Dona Castorina 110, 22460-320 Rio de Janeiro, RJ, Brazil. 
Phone:~+55\,21\,2529\,5070, Fax:~+55\,21\,2529\,5075, E-mail: alexei@impa.br}
}%
\affiliation{%
Institute of Mechanics, Lomonosov Moscow State University, Russia
}%

\date{\today}

\begin{abstract}
We consider self-similar solutions describing intermittent bursts in
shell models of turbulence, and study their relationship with
blowup phenomena in continuous hydrodynamic models. First, we show that these solutions 
are very close to self-similar solution for the Fourier transformed 
inviscid Burgers equation
corresponding to shock formation from smooth initial data. 
Then, the result is
generalized to hyperbolic conservation laws in one space dimension describing compressible
flows. It is shown that the renormalized wave profile tends to a universal
function, which is independent both of initial conditions and of a specific form
of the conservation law. This phenomenon can be viewed as a new manifestation of
the renormalization group theory. Finally, we discuss possibilities for application of the
developed theory for detecting and describing a blowup in
incompressible flows. 
\end{abstract}



\maketitle

\section{Introduction}

The theory of developed turbulence deals with unstable fluid flows at high
Reynolds numbers. This theory describes the energy transport from larger to 
smaller scales until the flow is smoothed at small scales due to
viscosity. A number of important phenomena are still not well understood and
described, e.g., the intermittency and anomalous
scaling in the inertial range~\cite{zakharov1992kolmogorov,frisch1999turbulence}. Intermittent
bursts, the spatially and temporarily localized fluctuations of the velocity
field, are principal ingredients of these phenomena. Such intermittent intense events
may be related to the blowup phenomenon in inviscid flow~\cite{holm2002transient,kerr2005velocity}, i.e., 
to a singularity developing in finite time from smooth initial conditions of finite energy in the case of vanishing viscosity. We know that, in
compressible inviscid flows, the blowup may lead to formation of a shock wave. 
The question of existence of the blowup in incompressible inviscid flows (incompressible Euler equations) 
represents a long standing and still controversial
problem, see, e.g., \cite{kerr1993evidence,hou2009blow,chae2008incompressible}.

This paper is focused on the study of scaling properties of blowups in
hydrodynamic flows. Our study is motivated by the success of the shell models of turbulence
to describe properties of the
developed turbulence in the inertial range~\cite{biferale2003shell}. The shell
models are simple ``toy'' models, where physical space is represented by a
discrete sequence of scales taken in geometric progression, and the velocity
field for each scale is represented by a single complex number. Turbulent bursts in
shell models are described by self-similar asymptotic
solutions, which were discovered and described numerically in
\cite{nakano1988,dombre1998intermittency,l2001outliers,l2002quasisolitons}. 
These solutions represent the blowup in inviscid shell models, i.e., 
a singularity developing in finite time, which we verify using the Beale-Kato-Majda type 
criterion proposed in~\cite{constantin2007regularity}.
Note that the self-similar solutions are important for explaining anomalous scaling 
of developed turbulence \cite{gilson1997towards,daumont2000instanton}. Analogous
self-similar solutions were found in more sophisticated cascade models
of incompressible Euler equations~\cite{uhlig1997singularities}. 

It is natural to relate self-similar blowup structures in discrete shell models
to blowup phenomena in realistic (continuous) hydrodynamic models. In this paper, we
accomplish this task for a blowup in compressible flows corresponding to the
well-known process of shock formation. First, we do that for the simplest nonlinear wave equation, the
inviscid Burgers equation, and then generalize the results for
hyperbolic conservation laws in one space dimension. A numerical method for detecting
self-similar structures in the Fourier space is developed by generalizing the
approach used in \cite{dombre1998intermittency} for shell models of turbulence.
This method works for systems with quadratic nonlinearities, thus, encompassing
a large variety of hydrodynamic models with quadratic convective terms. 

Self-similar structures in physical space are explained using the
renormalization group approach. We show that, near the blowup, 
there exists a limiting wave profile under renormalization of time, space and dependent variable. 
This limiting function (found in \cite{pomeau2008wave} for the case of Burgers equation) 
turns out to be universal, i.e., independent of
initial conditions as well as of a particular form of wave equation. 
Scaling properties of this universal solution agree with the Kolmogorov--Obukhov power law, 
which was not the case in the shell models. We compare the obtained results with more 
sophisticated universality phenomena like critical phenomena
of second-order phase transitions \cite{wilson1974renormalization,kadanoff2011}; 
see also \cite{eggers2009role} 
for the review of self-similar singularities in problems of fluid dynamics.  

The paper is organized as follows. In Section~2, we describe self-similar blowup
solutions in shell models of turbulence. Section~3 describes generalization of
these results for the Fourier transformed Burgers equation. In Sections~4 and 5
we return to the physical space and explain universality of renormalized blowup
solutions for compressible flows. Section~6 presents the renormalization group
approach. Section~7 discusses application of the developed theory for
incompressible flows. Section 8 summarizes the
contribution, and the Appendix contains some technical derivations.

\section{Self-similar structures in shell models of turbulence}

Shell models of turbulence are simple dynamical systems that demonstrate
important properties of developed turbulence in the Navier-Stokes equations at
high Reynolds numbers. In these models, the turbulent velocity field with wave
numbers $\mathbf{k}$ in the spherical shell $k_n < \|\mathbf{k}\| < k_{n+1}$ ($n
= 1,2,\ldots$) is represented by a single complex number $u_n$ called shell
velocity. The shell wave numbers are chosen as $k_n = k_0\lambda^n$ with the
most popular value $\lambda = 2$ for shell spacing. Equation for the shell
velocity is  
\begin{equation}
du_n/dt = Q_n[u]-\nu k_n^2u_n+f_n,
\label{eqS.0}
\end{equation}
where $Q_n[u]$ describes the quadratic nonlinearity, $\nu$ is the viscosity, and
$f_n$ is the external forcing applied to the first shells. It is assumed that
only several neighboring shells interact, as described by the nonlinear term of the form
\begin{equation}
Q_n[u] = i(k_{n+1}u_{n+2}u_{n+1}^*
-\varepsilon k_nu_{n+1}u_{n-1}^*
+(1-\varepsilon)k_{n-1}u_{n-1}u_{n-2})
\label{eqS.1}
\end{equation}
corresponding to the Sabra model~\cite{l1998improved}. For the GOY
model~\cite{gledzer1973system,ohkitani1989temporal}, the nonlinear term is
obtained by taking all shell speeds in (\ref{eqS.1}) with an asterisk and
substituting $(1-\varepsilon)$ by $(\varepsilon-1)$. The parameter is often chosen
to be $\varepsilon = 0.5$, in which case both the "energy" $\sum |u_n|^2$ and the
"helicity" $\sum (-1)^nk_n|u_n|^2$ are conserved in the system with no viscosity 
and no forcing, $\nu = f_n =
0$, reflecting basic properties of the   
Navier-Stokes equations.

\begin{figure}
\centering \includegraphics[width=0.6\textwidth]{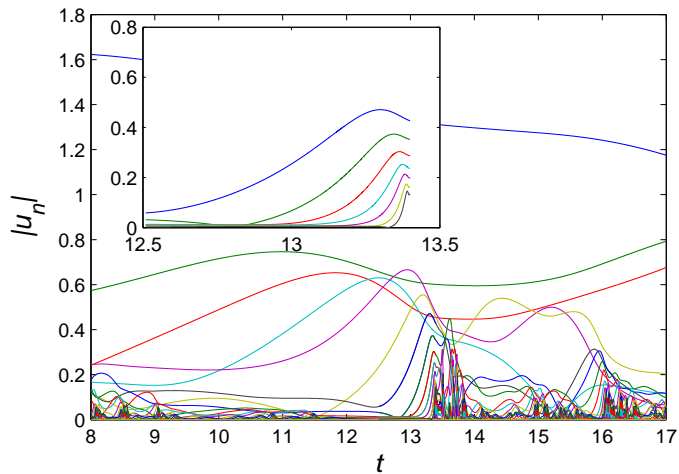}
\caption{Temporal dynamics of the Sabra shell model with $20$ shells, $\varepsilon = 0.5$, $k_0 =
2^{-4}$, $\nu = 2\times 10^{-6}$, $f_1 = 0.05$. Curves with smaller values of $|u_n|$ (in average) correspond to larger $n$. The inset shows the self-similar
structure of the shell velocities with $n = 7,\ldots,13$ for the largest turbulent
burst.}
\label{fig0}
\end{figure}

Fig.~\ref{fig0} shows typical behavior of the shell speeds. It is characterized by
a sequence of intermittent bursts, which have a self-similar structure shown in
the inset. In each burst, speed fluctuations propagate to smaller scales
(shells with larger $n$), until the viscosity becomes dominant
and smooths the oscillations. 
Self-similar structure of such bursts was described 
in~\cite{nakano1988,dombre1998intermittency,l2001outliers} 
by considering equations with vanishing viscosity $\nu = 0$. It was
shown numerically that the shell velocities behave
asymptotically for large $n$ as 
\begin{equation}
u_n(t) 
= -u_c(t_c-t)^{1/z-1}F\left(u_c(t_c-t)^{1/z}k_n\right), \quad
t < t_c.
\label{eqS.2}
\end{equation} 
We presented this expression in the different (equivalent) form, which is convenient for comparison with continuous models below. 
In (\ref{eqS.2}), $t_c$ is the final time and $u_c > 0$ is the scaling parameter. The function $F(k)$ is purely imaginary 
and defined up to scaling $F(k) \mapsto u_cF(u_ck)$, which reflects the 
symmetry of the shell model (\ref{eqS.0}), (\ref{eqS.1}) with $\nu = f_n = 0$ 
under the scaling $k_n \mapsto k_n/u_c$, $u_n \mapsto u_cu_n$. 
This function is universal for all the bursts. 
Expression (\ref{eqS.2}) describes identical structures, which appear in
subsequent shells renormalized by scaling of time and speeds. Numerical
computations provide the value 
\begin{equation}
z = 0.719 
\label{eqS.2b}
\end{equation}
for the GOY model with $\varepsilon = 0.5$~\cite{dombre1998intermittency}. For
the Sabra model, the best fit of rescaled velocity profiles for a specific
turbulent burst gave $z = 0.75\pm 0.02$~\cite{l2001outliers}. However, as we
will see below, the exponent $z$ has the same value (\ref{eqS.2b}) for both the GOY and Sabra
models.  This exponent is
different from $z = 2/3$ corresponding to the Kolmogorov--Obukhov law
$u_n \sim k_n^{-1/3}$ derived on dimensional grounds. This difference
has implications for explaining anomalous scaling in developed turbulence~\cite{l2001outliers}. 

At the final time $t_c$, the value of shell velocity $u_n$ is finite for any $n$ 
and, thus, no singularity is created for a particular shell. 
Despite of that we can show that expression (\ref{eqS.2}) describes a blowup, i.e., 
a singularity developing at $t_c$ for the infinite series of shells $1 \le n < \infty$. 
This can be done using the Beale-Kato-Majda type criterion of blowup in inviscid shell models, 
which requires~\cite{constantin2007regularity}
\begin{equation}
\lim_{t \to t_c^-}\int_0^{t} \sup_{n}k_n|u_n(s)|ds = \infty.
\label{eqBKM.1}
\end{equation}
Using (\ref{eqS.2}), we express 
\begin{equation}
k_n|u_n| = (t_c-t)^{-1}y|F(y)|,\quad 
y = u_c(t_c-t)^{1/z}k_n.
\label{eqBKM.1b}
\end{equation}
For small $t_c-t$ and $k_n = k_0\lambda^n$, one can always choose (large) $n$ 
such that $1 \le y \le \lambda$. Hence, 
\begin{equation}
\sup_{n}k_n|u_n(t)| \ge a(t_c-t)^{-1},\quad a = \min_{1 \le y \le \lambda} y|F(y)|,
\label{eqBKM.2}
\end{equation}
where $a$ is a positive constant.
This expression implies divergence of the integral in (\ref{eqBKM.1}). 

It is interesting to find a type of solution, which appears at $t = t_c$. Using (\ref{eqS.2}) and expressing $|u_n|$ similarly to (\ref{eqBKM.1b}), we find 
\begin{equation}
\lim_{t \to t_c^-}|u_n| = u_c^{z}k_n^{z-1}\lim_{y \to 0^+} y^{1-z}F(y).
\label{eqBKM.3}
\end{equation}
Since each $u_n$ is finite at $t_c$, 
the limit in (\ref{eqBKM.3}) is finite and $|u_n(t_c)| = bk_n^{z-1}$ for some positive constant $b$.
It follows that, at the blowup time $t_c$, the norm 
\begin{equation}
|u|_d = \left(\sum_{n = 1}^\infty k_n^{2d}|u_n|^2\right)^{1/2}
\label{eqBKM.4}
\end{equation}
is infinite for $d \ge 1-z = 0.281$ and finite for $d < 1-z$. 
Thus, we checked directly the blowup condition, 
which was introduced in \cite{constantin2007regularity} as the divergence 
of the norm with $d \ge 1$. Moreover, since $1-z < 1/3$, 
the (week) solution at $t \ge t_c$ does not satisfy the energy conservation 
criterion \cite{constantin2007regularity}. Hence, the blowup time $t_c$ 
may be viewed as a starting moment of the energy-cascade.

Expression (\ref{eqS.2}) can be explained using dynamical scaling of time and
shell speeds. This theory was proposed in~\cite{dombre1998intermittency}, 
and it is reviewed in the rest of this section. In our presentation, we changed the form of derivations and final self-similar expressions  facilitating development of the analogous theory for continuous hydrodynamic models. 

Let us write equation (\ref{eqS.0}) with $\nu = f_n = 0$ 
for the scaled speeds as
\begin{equation}
dv_n/dt = N_n[v],\quad
v_n = i k_n u_n, 
\label{eqS.3}
\end{equation}
with the nonlinearity
\begin{equation}
N_n[v] = -\lambda^{-2} v_{n+2}v_{n+1}^*
+\varepsilon v_{n+1}v_{n-1}^*
+(1-\varepsilon)\lambda^2 v_{n-1}v_{n-2}.
\label{eqS.4}
\end{equation}
For phenomena far from the first shells, i.e., for large $n$, we can consider
equation (\ref{eqS.4}) defined on the infinite lattice $n \in \mathbb{Z}$.  Then,
system (\ref{eqS.3}), (\ref{eqS.4}) is invariant under translations $n \mapsto
n+1$. For simplicity, we will assume that all $v_n$ are real, so that $u_n$ are
purely imaginary. In this case, the GOY and Sabra models are equivalent.

A translation-invariant system possessing a positive quadratic invariant can have traveling wave
solutions. Such a system can be obtained from (\ref{eqS.3}) by
introducing the dynamically scaled speeds $w_n(\tau)$ with the new time $\tau$ as  
\begin{equation}
t = t_c-\int_\tau^\infty \exp\left(-\int_0^{\tau'} A(\tau'') d\tau''
\right)d\tau',
\quad
v_n = \exp\left(\int_0^\tau A(\tau') d\tau'\right)w_n,
\label{eqS.6}
\end{equation}
where the blowup time $t_c$ corresponds to $\tau \to \infty$. It is
straightforward to check that
\begin{equation}
dw_n/d\tau = N_n[w]-Aw_n,
\label{eqS.5}
\end{equation}
where $N_n[w]$ has the form (\ref{eqS.4}) written in terms of $w_n$ instead of
$v_n$. Equation (\ref{eqS.5}) conserves the sum $\sum w_n^2$ if we take
\begin{equation}
A(\tau) = \sum_n w_nN_n[w]\Big/\sum_n w_n^2.
\label{eqS.5b}
\end{equation}
Note that transformation (\ref{eqS.6})--(\ref{eqS.5b}) is valid for any quadratic nonlinearity $N_n$.

\begin{figure}
\centering 
\includegraphics[width=0.32\textwidth]{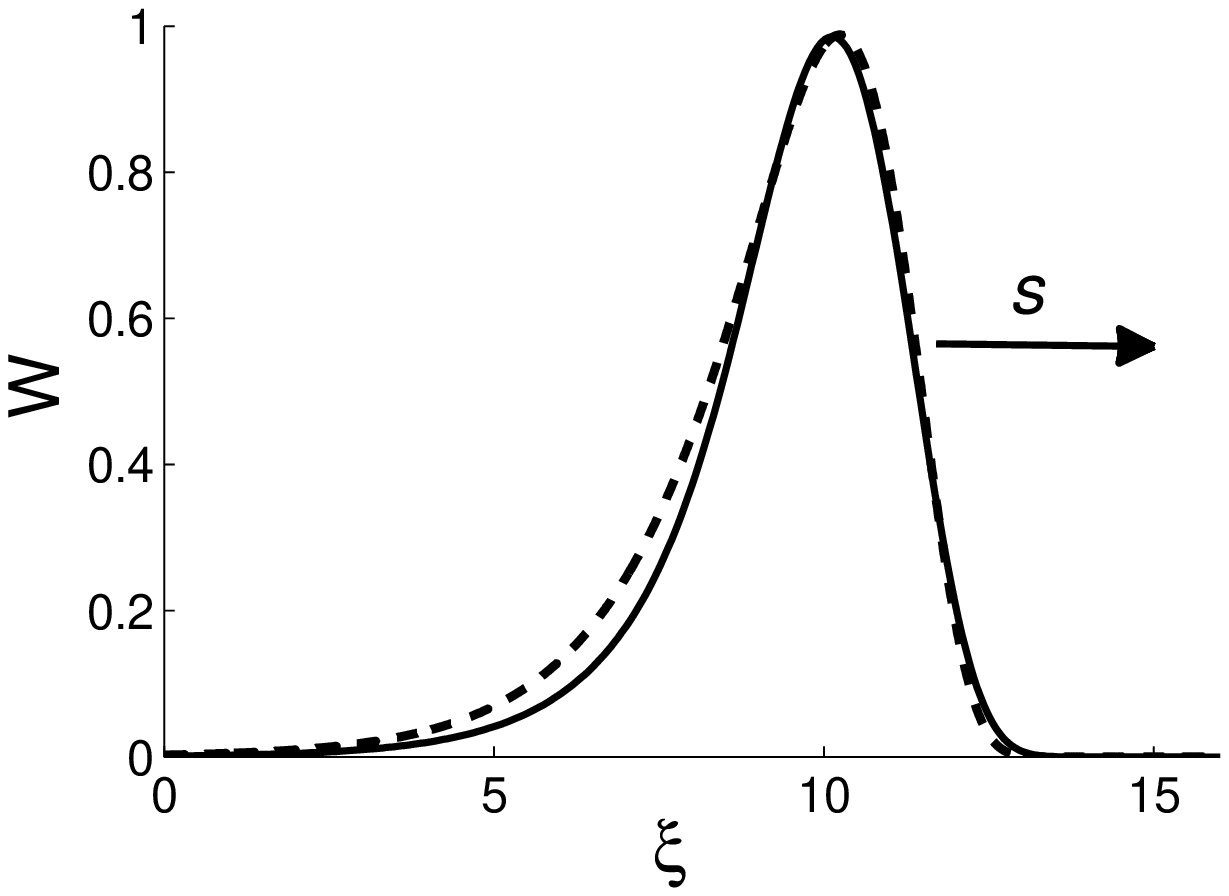}
\includegraphics[width=0.32\textwidth]{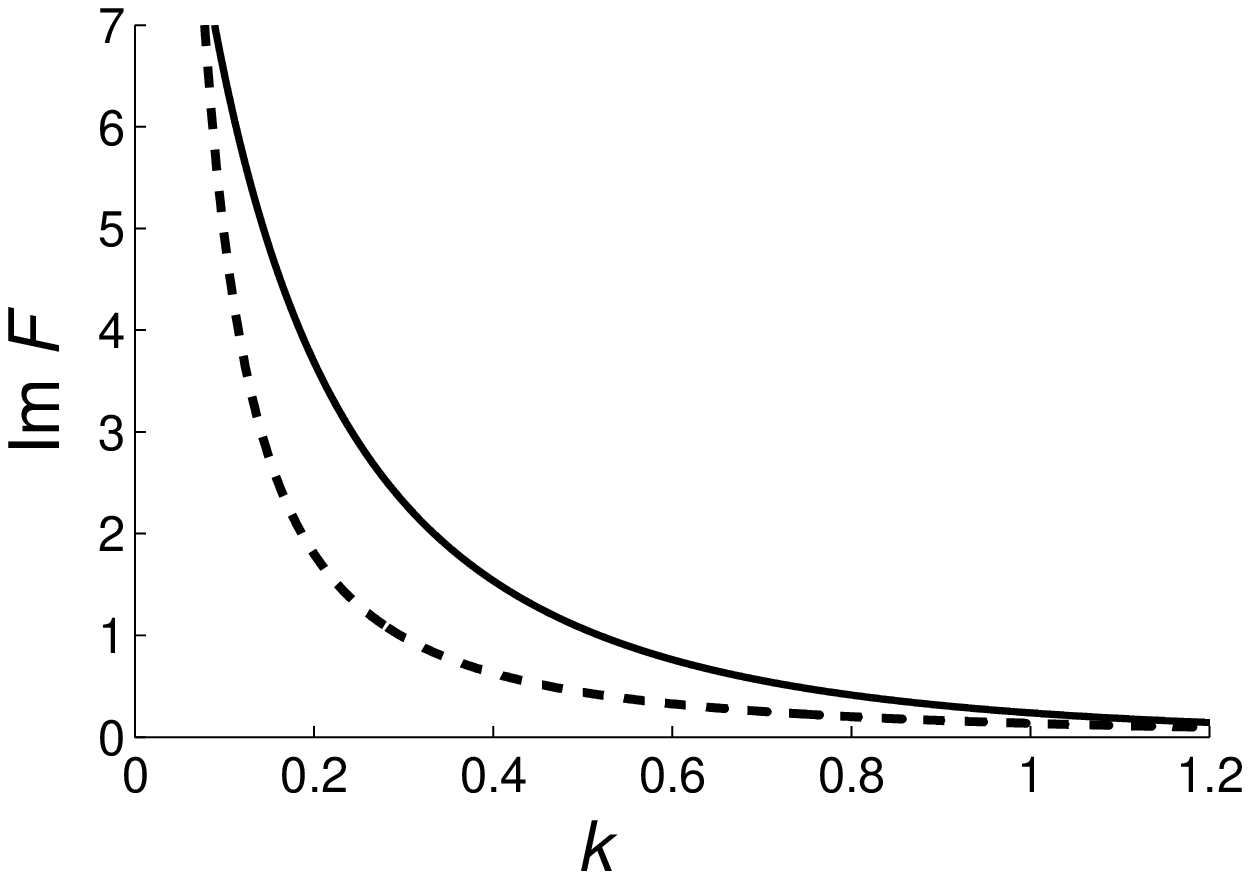}
\includegraphics[width=0.32\textwidth]{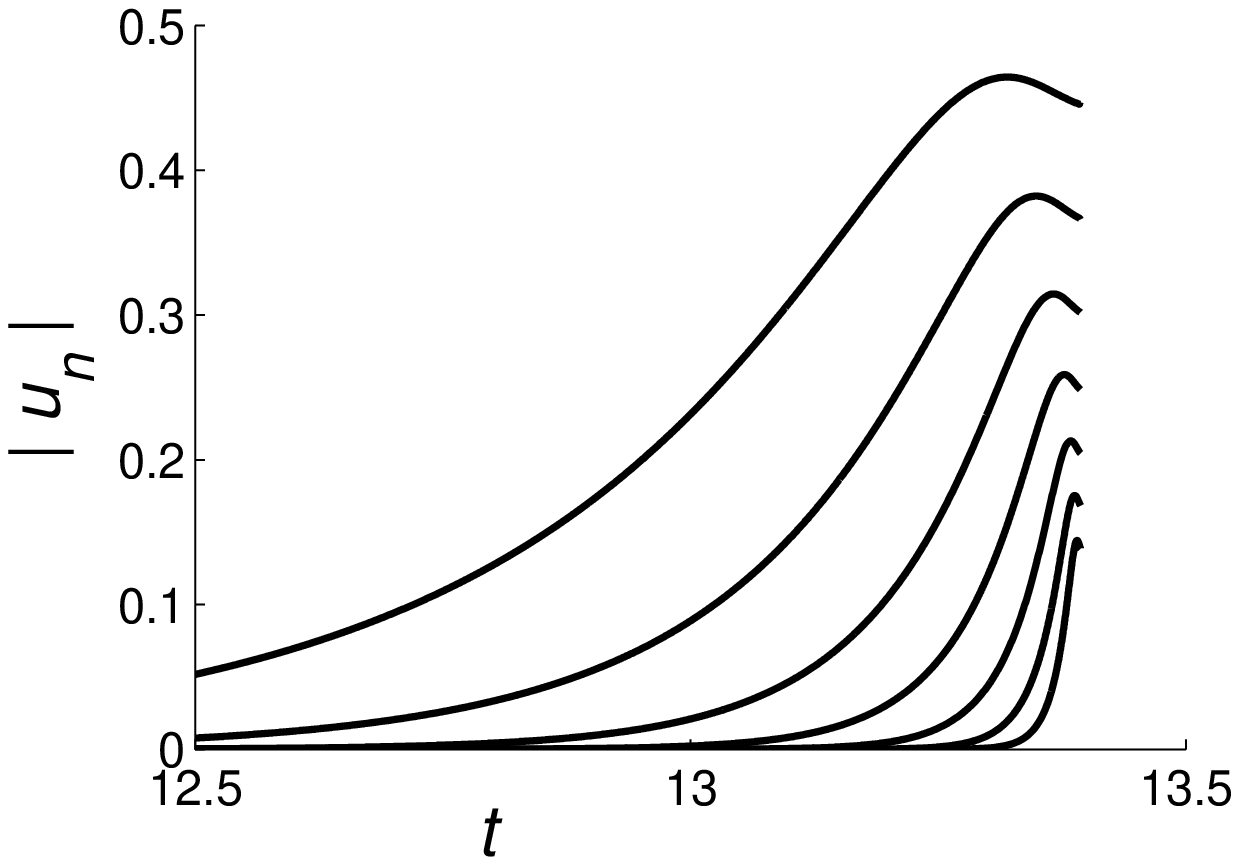}
\caption{Solid lines correspond to the shell model (Sabra or GOY) and dashed lines correspond
to the Burgers equation. Shown are the profiles $W(\xi)$ of solitary waves
(\ref{eqS.7}) and (\ref{eqB.7}), imaginary part of the universal functions $F(k)$
in (\ref{eqS.2}) and (\ref{eqB.5}), and the asymptotic form (\ref{eqS.2}) of shell
speeds $u_n(t)$ near the blowup with $t_c$ and $u_c$ adjusted to match the inset in
Fig.~\ref{fig0}.}
\label{fig3}
\end{figure}

Numerical computations~\cite{dombre1998intermittency} showed that equations
(\ref{eqS.5}), (\ref{eqS.5b}) have an asymptotically stable solution in the form of a solitary wave moving
toward large $n$ (small scales) with constant speed $s$, i.e., 
\begin{equation}
w_n(\tau) = W((n-s\tau)\log\lambda)
\label{eqS.7}
\end{equation}
for all $\tau$ and $n$; the function $W(\xi)$ vanishes as $\xi \to
\pm\infty$. The extra factor $\log\lambda$ is introduced here, which will be
convenient for comparison with continuous models below. Solution (\ref{eqS.7})
written in term of the original shell speeds $u_n(t)$ yields the self-similar
expression (\ref{eqS.2}) with the function
\begin{equation}
F(k) = ik^{z-1}\exp\left(\int_0^{\tau} A(\tau') d\tau'\right)W(-s\tau\log\lambda),\quad
z = \frac{1}{\log\lambda}\int_0^{1/s}A(\tau) d\tau,
\label{eqF1}
\end{equation} 
where $\tau$ is related to $t = t_c-k^z$ implicitly by (\ref{eqS.6}); 
see Appendix for the derivation. The function $W(\xi)$ of
the solitary wave (\ref{eqS.7}) and the corresponding purely imaginary function $F(k)$ 
given by (\ref{eqF1}) are presented in Fig~\ref{fig3}. The
graph on the right of Fig~\ref{fig3} shows shell velocities (\ref{eqS.2})
describing the asymptotic form of the blowup presented in the inset of
Fig.~\ref{fig0}. 

\section{Self-similarity of blowup for Burgers equation}

Self-similar structure (\ref{eqS.2}) of the blowup should have an analogue in
continuous hydrodynamic models. First, it is natural to analyze simple models
having well-known blowup structures. In this section, we start with the Burgers
equation $u_t +uu_x = \nu u_{xx}+f$, where $\nu$ is the small viscosity, $f$ is
the external forcing, and subscripts denote partial derivatives. The Burgers
equation models various phenomena in compressible fluids, multiphase flows in porous
media, etc. For the Fourier transformed wave profile $u(k) = (2\pi)^{-1}\int
u(x)e^{-ikx}dx$, this equation takes the form 
\begin{equation}
\frac{\partial u}{\partial t} 
= -i\int_{-\infty}^\infty 
(k-p)u(p)u(k-p)dp-k^2\nu u+f.
\label{eqB.1}
\end{equation} 
For real functions $u(x)$ decaying sufficiently fast at infinity, the Fourier transform $u(k)$ is regular and
$u(-k) = u^*(k)$. Thus, we can consider only positive $k$ from now on. 
Additionally, we assume that $u(k)$ is purely imaginary, 
which means that the function $u(-x) = -u(x)$ is odd in  physical space 
(we return to the general case in the
next section).

In order to study a blowup, we take $\nu = f = 0$. 
Then, there is a global invariant $\int u^2dx$,
which corresponds to $\int |u|^2dk$ in the Fourier space.
(Also, one can treat $\int uu_xdx = 0$ as a second global invariant analogous to helicity.)
Hence, equation (\ref{eqB.1}) can be related to the shell model (\ref{eqS.0}),
(\ref{eqS.1}) with the same reasoning as for the
Navier-Stokes equations. 
The shell wave numbers $k_n = k_0\lambda^n$
growing in geometric progression induce the expression $k
= e^\xi$ in the continuous formulation. For the logarithmic coordinate $\xi$, we write
(\ref{eqB.1}) with $\nu = f = 0$ as
\begin{equation}
\partial v/\partial t = N[v],\quad
v(\xi) = -ik^2u(k), 
\label{eqB.2}
\end{equation}
with the real function $v(\xi)$ and the nonlinear term
\begin{equation}
\begin{array}{c}
\displaystyle
N[v] = \sum_{\sigma = \pm}\int_{-\infty}^\infty v(\eta)v(\zeta)
\frac{k^2}{pq}d\eta,\\[15pt]
\displaystyle
k = e^{\xi},\ p = \sigma e^{\eta}, \ q = e^\zeta = |k-p|,
\end{array}
\label{eqB.3}
\end{equation}
where we used the relations 
\begin{equation}
\begin{array}{c}
u(p)dp = u(p)|p|d\eta = u(|p|)pd\eta = (i/p)v(\eta)d\eta, \\[7pt]
(k-p)u(k-p) = qu(q) = (i/q)v(\zeta)
\end{array}
\label{eqB.3b}
\end{equation}
following from $u(-k) = u^*(k) = -u(k)$ for the purely imaginary $u(k)$.
One can see the analogy of equations (\ref{eqB.2}), (\ref{eqB.3}) with the discrete system
(\ref{eqS.3}), (\ref{eqS.4}). In particular, $N[v]$ is quadratic with respect to $v$ and invariant
under translations $\xi \mapsto \xi+\Delta\xi$. 

We can write (\ref{eqB.2}), (\ref{eqB.3}) in the form similar to (\ref{eqS.5}), (\ref{eqS.5b}) as
\begin{equation}
dw/d\tau = N[w]-Aw,\quad
A = \int_{-\infty}^\infty wN[w]d\xi
\left(\int_{-\infty}^\infty w^2d\xi\right)^{-1}.
\label{eqB.6}
\end{equation}
The real function $w(\tau,\xi)$ is related to $v(t,\xi)$ by expressions
(\ref{eqS.6}), where the subscript $n$ must be dropped. 
One can check that equation (\ref{eqB.6}) conserves the integral $\int w^2d\xi$.
Hence, it allows traveling wave solutions. 

Numerical integration of (\ref{eqB.6}) shows that there exists a stable
solution 
\begin{equation}
w(\tau,\xi) = W(\xi-s\tau),
\label{eqB.7}
\end{equation}
representing a solitary wave moving with constant speed $s$ in positive
direction of $\xi$-axis, i.e., toward small scales $x \sim 1/k = e^{-\xi}$. The profile $W(\xi)$
of this wave is shown in Fig.~\ref{fig3} by the dashed line, and one can notice that it is very close
to the corresponding profile (\ref{eqS.7}) obtained for the shell model of turbulence and shown by the solid line. 
Derivations given in the Appendix lead to the self-similar solution
analogous to (\ref{eqS.2}) of the form
\begin{equation}
u(t,k) 
= u_c^2(t_c-t)^{2/z-1}F\left(u_c(t_c-t)^{1/z}k\right), \quad
t < t_c,
\label{eqB.5}
\end{equation}
with arbitrary $u_c > 0$ and 
\begin{equation}
F(k) = \frac{i}{Ak^2}W\left(\log k+\frac{\log A}{z}\right),
\quad z = \frac{A}{s},
\label{eqB.5b}
\end{equation}
where the constant $A$ is given by (\ref{eqB.6}).
The exact Kolmogorov value of the exponent 
\begin{equation}
z = 2/3 
\label{eqB.5c}
\end{equation}
is found in this case. A graph of the purely imaginary function $F(k)$ is 
presented in Fig.~\ref{fig3} (dashed line). 

We see that the traveling wave solution (\ref{eqS.7}) for 
the shell model of turbulence has a direct analog (\ref{eqB.7}) for the inviscid Burgers equation, with the corresponding self-similar solutions (\ref{eqS.2}), (\ref{eqF1}) and (\ref{eqB.5}), (\ref{eqB.5b}). The result (\ref{eqB.5c}) suggests that the anomalous value of $z$ in
(\ref{eqS.2b}) may be attributed to the approximate nature of the shell models.
Physical interpretation of the obtained self-similar solutions is given in the next section.

\section{Self-similar blowup structure in physical space}

The self-similar blowup solution just described must correspond to the blowup (formation
of a shock wave) in the inviscid Burgers equation 
\begin{equation}
u_t +uu_x = 0. 
\label{H.1}
\end{equation}
The solution $u(t,x)$ constructed by the method of characteristics has the
implicit form
\begin{equation}
x = x_0+u_0(x_0)(t-t_0),\quad
u = u_0(x_0),
\label{H.2}
\end{equation}
where $u(t_0,x) = u_0(x)$ is a smooth initial condition and $x_0$ is an
auxiliary variable.  
For spatial derivative of $u(t,x)$, we have
\begin{equation}
\frac{\partial u}{\partial x}
= \frac{\partial u/\partial x_0}{\partial x/\partial x_0} 
= \frac{u_0'(x_0)}{1+u_0'(x_0)(t-t_0)}.
\label{H.3}
\end{equation}
The denominator vanishes at $t = t_0-1/u_0'(x_0)$.
This yields the well-known result that the classical solution blows up along the
characteristic with the minimum negative value of $u_0'(x_0)$, followed by the
formation of a shock wave. 

Let us choose the origin of time and space  so that the blowup singularity appears
at $t = x = 0$, and consider a smooth solution in the interval $t_0 \le t <
0$. Also, we can take $u = 0$ at the singularity, which can be achieved by the
transformation $x \mapsto x-u_0(0)t$ and $u \mapsto u+u_0(0)$, which
leaves (\ref{H.1}) invariant. In this case, the initial condition $u_0(x)$
satisfies the conditions
\begin{equation}
u_0(0) = 0,\quad t_0 = 1/u_0'(0) < 0,\quad u_0''(0) = 0,\quad u_0'''(0) > 0,
\label{H.4}
\end{equation}
which ensure that $u_0'(x)$ has a negative minimum at the origin.
Note that, for the odd functions $u(x)$ considered in the previous
section, the conditions $u_0(0) = u_0''(0) = 0$ are satisfied automatically.

For small $x_0$, we use (\ref{H.4}) to expand 
\begin{equation}
u_0(x_0)t_0 = x_0+\frac{u_0'''(0)}{6u'(0)}x_0^3+o(x_0^3).
\label{H.5a}
\end{equation}
Using this expansion in (\ref{H.2}) yields
\begin{equation}
x = ut-\frac{u_0'''(0)}{6u'(0)}x_0^3+o(x_0^3),\quad
u = u'_0(0)x_0+o(x_0).
\label{H.5}
\end{equation}
Equivalently,
\begin{equation}
x = ut-cu^3+o(u^3),\quad
c = \frac{u_0'''(0)}{6(u'(0))^4} > 0.
\label{H.6}
\end{equation}

Consider now the renormalization operator $\mathcal{G}_{\lambda}$ acting as
\begin{equation}
u_\lambda(t,x) = \mathcal{G}_{\lambda}u(t,x) 
\equiv \lambda^{1/3}u(\lambda^{-2/3}t,\,\lambda^{-1} x).
\label{H.7}
\end{equation}
It is easy to see that $u_\lambda$ is a new (renormalized) solution of (\ref{H.1}).
Thus, (\ref{H.7}) represents a symmetry of (\ref{H.1}). Multiplying
both sides of (\ref{H.6}) by $\lambda$ and making the substitution $t \mapsto
\lambda^{-2/3}t$, $x \mapsto \lambda^{-1}x$ yields the equation for
$u_\lambda(t,x)$ as
\begin{equation}
x = u_\lambda t-cu_\lambda^3+\lambda\,
o\!\left((\lambda^{-1/3}u_\lambda)^3\right).
\label{H.8}
\end{equation}
The last (correction) term contains powers $\lambda^{1-n/3}$ with $n
> 3$. Hence, it vanishes in the limit of large $\lambda$,
and (\ref{H.8}) takes the exact form \cite{pomeau2008wave,eggers2009role}
\begin{equation}
x = u_\infty t-cu_\infty^3
\label{H.10}
\end{equation}
for the limiting function
\begin{equation}
u_\infty(t,x) = \lim_{\lambda \to \infty}u_\lambda(t,x).
\label{H.9}
\end{equation}
Clearly, the function $u_\infty(t,x)$ is self-similar, 
\begin{equation}
\mathcal{G}_\lambda u_\infty = u_\infty.
\label{R.1}
\end{equation}
We write solution of (\ref{H.10}) as
\begin{equation}
u_\infty(t,x) = u_c(-t)^{1/2}F(u_c^{-1}(-t)^{-3/2}x),\quad u_c = c^{-1/2}, \quad
t < 0,
\label{R.2}
\end{equation}
where $F(x)$ is defined uniquely by the equation 
\begin{equation}
x = -F-F^3, 
\label{eqR.2b}
\end{equation}
as one can check by substituting (\ref{R.2}) into (\ref{H.10}). 
Note that the factor $u_c$ reflects the scaling
symmetry of the Burgers equation (\ref{H.1}). 

We see that, near the blowup, the renormalized solution $u_\lambda(t,x)$ is described asymptotically
by the function $F(x)$ in (\ref{R.2}). Fourier transform of (\ref{R.2})
yields the self-similar expression (\ref{eqB.5}) with $z = 2/3$ and $t_c = 0$,
where $u(t,k)$ is the Fourier transformed function $u_\infty(t,x)$. This
explains self-similarity of blowup in Fourier transformed Burgers equation with
the exact Kolmogorov value $z = 2/3$ of the scaling exponent. Also, the function $F(k)$
describing self-similarity in the Fourier space is the Fourier transform of $F(x)$ defined in (\ref{eqR.2b}). This completes analytical
description of the self-similar blowup structure for the inviscid Burgers
equation.  

For connection with the classical theory, we can consider the limit $t \to 0$ with 
fixed $x \ne 0$, when the argument of the function $F$ in (\ref{R.2}) becomes large. 
Equation (\ref{eqR.2b}) yields $F(x) \approx -x^{1/3}$ for the large argument $x \gg 1$. Using this asymptotic 
formula in (\ref{R.2}), we obtain the cubic root relation 
$u_\infty \approx -u_c^{2/3}x^{1/3}$ at the singular point, 
which is well-known. Note that (\ref{R.2}) represents a much stronger result, 
since it suggests that the whole renormalized profile 
(with derivatives of any order) becomes universal. 

\begin{figure*}
\centering \includegraphics[width=0.49\textwidth]{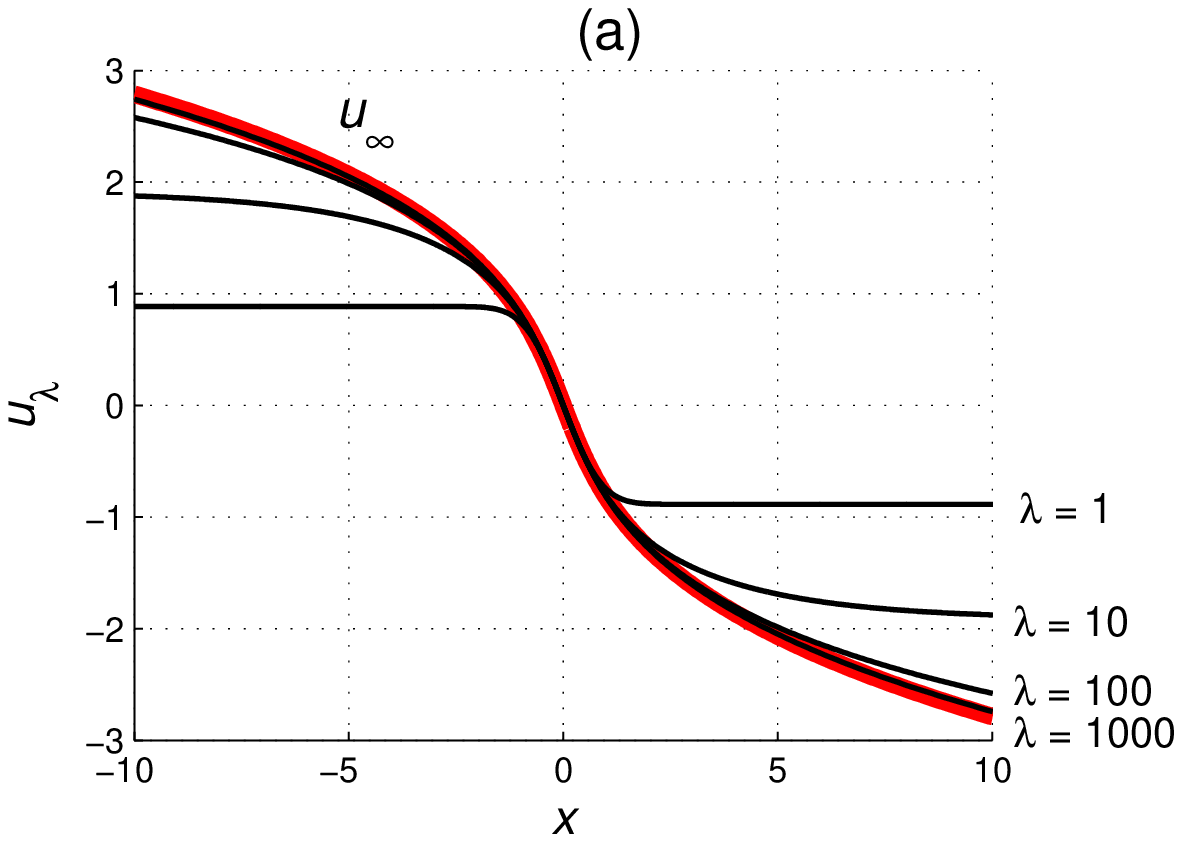}
\centering \includegraphics[width=0.49\textwidth]{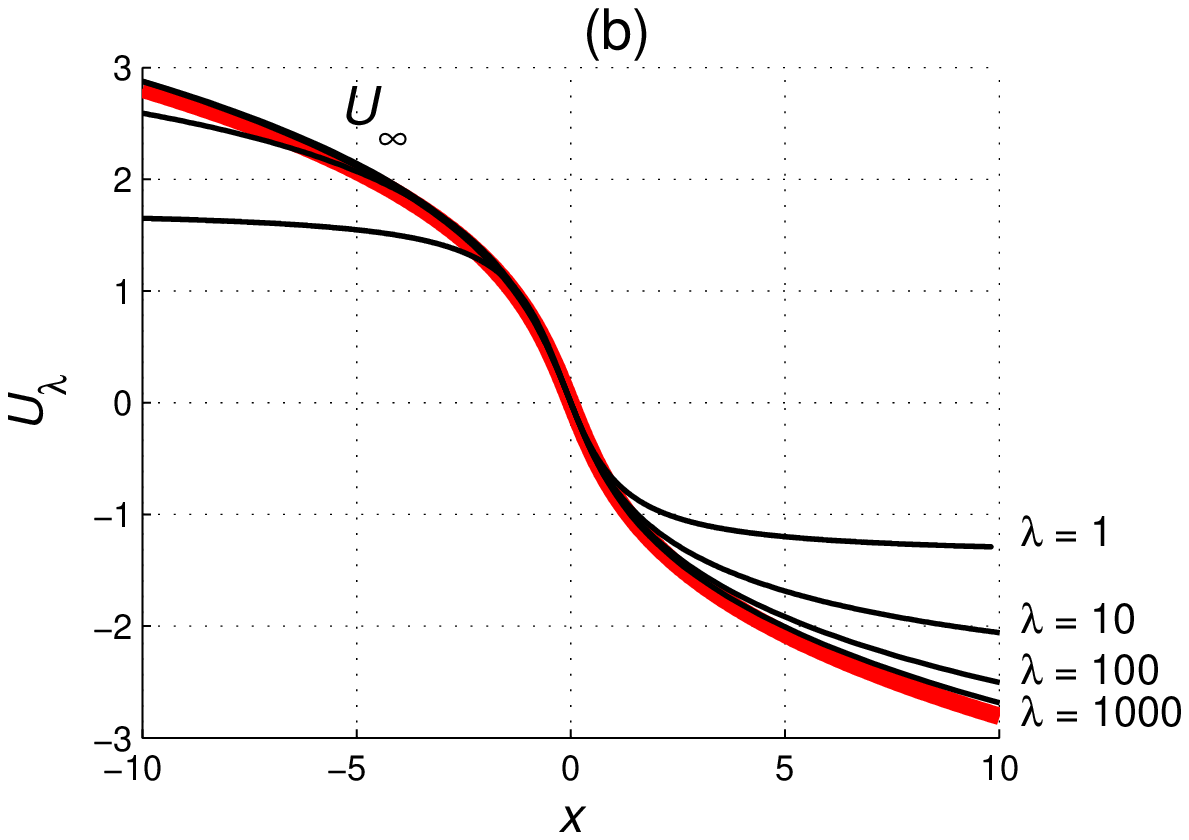}
\caption{(a) Convergence of the renormalized profiles $u_\lambda(t_0,x)$ (black
curves) to the universal form $u_\infty(t_0,x)$ (bold red curve) given by (\ref{R.2}), (\ref{eqR.2b}) for the inviscid
Burgers equation (\ref{H.1}). The initial condition is $u_0(x) =
-(\sqrt{\pi}/2)\textrm{erf}\,x$ at $t_0 = -1$, and $\lambda =
1,\,10,\,10^2,\,10^3$. (b) Similar convergence result for blowup of
a simple wave in ideal polytropic gas. The bold red curve represents the 
universal limiting function $U_\infty(t_0,x)$ 
given by (\ref{C.3}), (\ref{R.2}), (\ref{eqR.2b}) 
and back curves are the renormalized profiles
$U_\lambda(t_0,x) = \mathcal{G}_\lambda U(t_0,x)$ corresponding to 
the gas density variation $U$.} 
\label{fig1}
\end{figure*}

Fig.~\ref{fig1}a presents a 
numerical example of convergence of the renormalized solution $u_\lambda(t_0,x)$ 
to the self-similar form (\ref{R.2}) with the increase of 
$\lambda$ for $t_0 = -1$ 
and the initial condition $u_0(x) = -(\sqrt{\pi}/2)\textrm{erf}\,x$. 
We also confirmed
numerically that $F(k)$ is the Fourier transform of $F(x)$. 

We conclude that the self-similar 
solution (\ref{eqB.5}) obtained by the Dombre--Gilson 
method successively capture the universality of renormalized profiles 
near the blowup in the inviscid Burgers equation.
This universality result turns out 
to be much more general, as we will see below.

\section{Universality of blowup in compressible flows}

Compressible (or multi-phase) inviscid flows in one space dimension 
are modeled by hyperbolic systems of
conservation laws. In scalar case, $U \in \mathbb{R}$, we consider the equation
\begin{equation}
\frac{\partial U}{\partial t}
+\frac{\partial f(U)}{\partial x} = 0
\label{C.1}
\end{equation}
with a smooth flux function $f(U)$ and initial condition $U_0(x)$ at $t =
t_0$. A classical solution of equation (\ref{C.1}) has the form 
\begin{equation}
x = x_0+f'(U_0(x_0))(t-t_0),\quad
U = U_0(x_0),
\label{C.2b}
\end{equation}
which reduces to solution (\ref{H.2})
of the Burgers equation by the substitution $u = f'(U)$. We assume that the
coordinates for time and space are chosen such that (\ref{H.4}) holds for
$u_0(x) = f'(U_0(x))$, which implies the blowup at $t = x = u = 0$. 
Assuming that $u = f'(U)$ is locally invertible as $U = g(u)$ with
$g(0) = 0$ and $g'(0) = 1/f''(0)$, we have the blowup at $U = 0$ for the 
original state variable. We will use the relation
\begin{equation}
\mathcal{G}_{\lambda}u^n(t,x)
= \lambda^{1/3}u^n(\lambda^{-2/3}t,\,\lambda^{-1} x) 
= \lambda^{(1-n)/3}u_{\lambda}^n(t,x)
\label{C.2}
\end{equation}
with the renormalization operator $\mathcal{G}_\lambda$ given in (\ref{H.7}).
Expanding $g(u)$ in Taylor series 
\begin{equation}
g(u) = g(0)+g'(0)u+\frac{1}{2}g''(0)u^2+\cdots 
= \frac{u}{f''(0)}+\frac{1}{2}g''(0)u^2+\cdots,
\label{C.2bb}
\end{equation}
we find
\begin{equation}
U_\infty(t,x) = \lim_{\lambda\to\infty}\mathcal{G}_{\lambda}U(t,x) 
= \lim_{\lambda\to\infty}\mathcal{G}_{\lambda}g(u(t,x)) 
= u_\infty(t,x)/f''(0),
\label{C.3}
\end{equation}
where all terms $\mathcal{G}_{\lambda}u^n$ with $n > 1$ vanish 
for $\lambda \to \infty$ due to (\ref{C.2}).

Formulas (\ref{H.7}), (\ref{C.3}) with the function $u_\infty(t,x)$ given
by (\ref{R.2}), (\ref{eqR.2b}) demonstrate strong universal character of the blowup in scalar 1D
conservation laws. We see that, when the singular point is approached, the
renormalized wave profile $U(t,x)$ takes the universal form independent both of
the initial condition $U_0(x)$ and of the flux function $f(U)$. In particular, derivatives of
all orders turn out to be universal for the renormalized solution near the
singular point. Note that expressions (\ref{H.6})--(\ref{H.9}) relate the blowup with a cusp
catastrophe in the space $(t,x,u)$, see \cite{arnoldcatastrophe}. Therefore,
universality of the blowup just described represents a property of this catastrophe.

As an example, let us consider formation of a shock in a simple wave
solution for one-dimensional flow of ideal polytropic gas. The gas density
$\rho(t,x)$ in this wave is described implicitly by
\begin{equation}
\frac{x-x_0}{t-t_0} =
\frac{\gamma+1}{\gamma-1}\sqrt{A\gamma}\rho_0^{(\gamma-1)/2}(x_0),\quad 
\rho = \rho_0(x_0),
\label{G.1}
\end{equation}
where $\rho_0(x)$ is the initial condition at $t = t_0$, 
see, e.g., \cite{courant1977supersonic}. We will use the values $\gamma = 5/3$,
$A = 3/5$, and $\rho_0(x) = 2-\arctan x$. 
Then expressions (\ref{G.1}) take the form
\begin{equation}
\begin{array}{l}
x = x_0+4(2-\arctan x_0)^{1/3}(t-t_0),\\[5pt]
\rho = 2-\arctan x_0,
\end{array}
\label{G.1b}
\end{equation}
which can be written in the form (\ref{C.2b}) for the density variation in
moving frame
\begin{equation}
U(t,x) = \rho(t,x+x_1+v_1(t-t_0))-\rho_1
\label{G.2c}
\end{equation}
with
\begin{equation}
\begin{array}{c}
f'(U) = 4(U+\rho_1)^{1/3}-v_1,\\[3pt]
U_0(x) = 2-\arctan (x+x_1)-\rho_1.
\end{array}
\label{G.2}
\end{equation}
The quantities $t_0$, $x_1$, $\rho_1$ and $v_1$ 
are chosen such that $f'(0) = 0$ and the function 
\begin{equation}
u_0(x) = f'(U_0(x)) = 4(2-\arctan(x+x_1))^{1/3}-v_1
\label{G.2d}
\end{equation}
satisfies conditions (\ref{H.4}) corresponding to blowup at $t = x = 0$ with $U
= 0$. The values of these quantities are
\begin{equation}
t_0 = -1.155,\ 
x_1 = 0.183,\ 
\rho_1 = 1.818,\ 
v_1 = 4.882.
\label{G.2e}
\end{equation} 
Convergence (\ref{C.3}) of the renormalized profile of gas density variation 
$U_\lambda(t,x) =
\mathcal{G}_\lambda U(t,x)$ to the universal limiting form $U_\infty$ is demonstrated in
Fig.~\ref{fig1}b. 

\section{Renormalization group}

The scaling universality of blowup can also be explained using the
renormalization group approach. For this purpose, we consider
$\mathcal{G}_\lambda$ in (\ref{H.7}) as an operator acting in the space of
solutions of the Burgers equation (\ref{H.1}) with initial conditions satisfying
(\ref{H.4}). Then the system evolution can be seen as the action of
$\mathcal{G}_\lambda$ combined with renormalization of space $x$ and state $u$. The
operator $\mathcal{G}_\lambda$ defines a differentiable one-parameter group with the property
$\mathcal{G}_{\lambda_1}\mathcal{G}_{\lambda_2} =
\mathcal{G}_{\lambda_1+\lambda_2}$. The self-similar solution $u_\infty(t,x)$ in
(\ref{R.2}), (\ref{eqR.2b}) represents a stationary point (\ref{R.1}) of the renormalization
group operator. This stationary point (more precisely,
a set of stationary points $u_\infty(t,x)$ determined up to a
scaling constant $u_c$) is asymptotically stable in the sense of Lyapunov for
$\lambda \to \infty$ considered as ``time" \cite{eggers2009role}. 
This stability condition implies the
universal limit (\ref{H.9}). 

In the case of a general conservation law (\ref{C.1}), the renormalized function
$U_\lambda(t,x) = \mathcal{G}_\lambda U(t,x)$ is a solution for a conservation
law with the renormalized flux function  $f_\lambda(U) =
\lambda^{2/3}f(\lambda^{-1/3}U)$. Thus, $\mathcal{G}_\lambda$ defines a one-parameter 
group acting in the functional space of solutions and fluxes as $(U,f) \mapsto
(U_\lambda,f_\lambda)$. The universality (\ref{C.3}) of
the blowup is explained by the fact that $\mathcal{G}_\lambda$ has the
asymptotically stable stationary point $(U,f) = (u_\infty(t,x),U^2/2)$
corresponding to solution (\ref{R.2}) of Burgers equation
(\ref{H.1}).  

The role of the operator $\mathcal{G}_\lambda$ is similar to renormalization group 
operators in other physical theories. For the inner scale, $x
\sim ut \sim (-t)^{3/2}$ with small $t$, the dynamics is governed by the
universal function, which is a stationary point of $\mathcal{G}_\lambda$. This
is analogous, e.g., to the stationary point of the renormalization group
operator, which determines critical exponents in second-order phase
transitions~\cite{wilson1974renormalization}. At larger spatial scales, there is
no universality and the solution depends on the initial condition $u_0(x)$ as
well as on the flux function $f(u)$. This is analogous, in turn, to the
phenomenological Landau theory of second-order phase transitions valid at larger
(though still small) deviations of temperature from a critical
value~\cite{landau1980statistical}.

Up to now, we considered compressible flows, where the blowup corresponds to the
well-known phenomenon of shock formation. Could the presented 
approach be used for the study of incompressible flows? 
We address this question in the next section.
Such an application is strongly motivated by the conclusion, 
which was made in \cite{kerr2005velocity}
based on the scaling analysis of a blowup in numerical solutions of incompressible Euler equations: "it could be that the antiparallel configuration
is universal as $t \to T$ and $r < R$. For large $r$ there is a helical regime
that might not be universal..." This hypothesis is exactly what the
renormalization group theory described above suggests for the blowup solution. 

\section{Dombre--Gilson scheme for incompressible Euler equations}

The procedure described in Section 3 works for evolutionary systems 
with quadratic nonlinearities written in the Fourier space. This covers a large
class of hydrodynamic models with quadratic convection terms. Let us consider the
incompressible Euler equations written for the Fourier transformed velocity
$\mathbf{u}(t,\mathbf{k})$ as 
\begin{equation}
\partial\mathbf{u}/\partial t
= -i\mathbf{P}\int d^3p\,
(\mathbf{u}(\mathbf{p}) \cdot \mathbf{k})\,
\mathbf{u}(\mathbf{k}-\mathbf{p})
\label{eqE.1}
\end{equation}
with the incompressibility 
condition $\mathbf{k}\cdot \mathbf{u} = 0$.
Here $\mathbf{P} = \mathbf{I}-\mathbf{k}\mathbf{k}^T/k^2$ with $k = \|\mathbf{k}\|$
is the projector onto transverse vector fields $\mathbf{k}\cdot \mathbf{u} = 0$, and the incompressibility condition is used to remove the pressure and $\mathbf{u}(\mathbf{p}) \cdot \mathbf{p}$ terms. 
Considering logarithmic spherical coordinates $\xi = \log k$, $\mathbf{o}_k =
\mathbf{k}/k$, we write equation for the scaled speed in the form similar to
(\ref{eqB.2}) as
\begin{equation}
\partial\mathbf{v}/\partial t = N[\mathbf{v}],\quad
\mathbf{v}(\xi,\mathbf{o}_k) = -ik^4\mathbf{u}(\mathbf{k}),
\label{eqE.2}
\end{equation}
\begin{equation}
N[\mathbf{v}] 
= \mathbf{P}\int d^2\mathbf{o}_pd\eta\,
\frac{k^5}{pq^4}\,(\mathbf{v}(\mathbf{p}) \cdot \mathbf{o}_k)\,
\mathbf{v}(\mathbf{k}-\mathbf{p})
,
\label{eqE.3}
\end{equation}
where $p = \|\mathbf{p}\|$, $q = \|\mathbf{k}-\mathbf{p}\|$, $\eta = \log p$, $\mathbf{o}_p = \mathbf{p}/p$ 
and the integration is carried over $-\infty < \eta < \infty$ and 
the unit sphere $\mathbf{o}_p \in S^2$.
One can check that this equation is invariant under translations $\xi \mapsto
\xi+\Delta\xi$ of the vector field $\mathbf{v}(\xi,\mathbf{o}_k)$, 
which represent scaling in the Fourier space, 
$k \mapsto e^{\Delta\xi} k$. We assume that $\mathbf{v}$ is a real vector field 
corresponding to a flow with the symmetry 
$\mathbf{u}(-\mathbf{x}) = -\mathbf{u}(\mathbf{x})$.
Since the nonlinearity $N[\mathbf{v}]$ is quadratic, 
we can transform (\ref{eqE.3}) to the
form similar to (\ref{eqB.6}) as
\begin{equation}
\partial\mathbf{w}/\partial\tau
= N[\mathbf{w}]-A\mathbf{w},
\quad
A = 
\left(\int d^2\mathbf{o}_kd\xi\,\|\mathbf{w}\|^2
\right)^{-1}
\int d^2\mathbf{o}_kd\xi\,\mathbf{w}\cdot N[\mathbf{w}],
\label{eqE.4}
\end{equation}
where $\tau$ and $\mathbf{w}$ are given by (\ref{eqS.6}) with $v_n$ and $w_n$
substituted by $\mathbf{v}$ and $\mathbf{w}$. System (\ref{eqE.4}) is
translational-invariant along direction $\xi$ and preserves the norm $\int
d^2\mathbf{o}_kd\xi\,\|\mathbf{w}\|^2$.
Hence, it admits traveling wave solutions of the form 
$\mathbf{w}(\tau,\xi,\mathbf{o}_k) = \mathbf{W}(\xi-s\tau,\mathbf{o}_k)$. 
Such solutions determine a self-similar blowup described in physical space 
by an expression analogous to (\ref{eqB.5}), which was considered first by 
Leray~\cite{leray1934mouvement} in this context.

Numerical computations for system (\ref{eqE.4}) were performed, 
where the vector spherical harmonics expansion with coefficients 
depending on $\xi$ was used for the function $\mathbf{w}(\xi,\mathbf{o}_k)$. 
Such expansion has advantages for numerical computation of the operator $N[\mathbf{w}]$.
Details of the numerical method and its realization will be 
published elsewhere. Here we comment only on the conclusions 
related to the presented approach. First of all, no traveling 
wave solutions were found. On the other hand, we observed that
at least two characteristic scales describe formation of a singularity, since the solution was concentrated in a small 
region on the unit sphere $\mathbf{o}_k \in S^2$ with increasing time $\tau$ 
and wavenumber $k$. This conclusion agrees with the numerical analysis
in \cite{kerr2005velocity} suggesting the two scales, $r \sim (t_c-t)$ and
$(t_c-t)^{1/2}$, where $t_c$ is the time of anticipated collapse. 

Note that equations (\ref{eqE.4}) account for self-similar
solutions characterized by a single scale $x \sim k^{-1} = e^{-\xi}$. 
Such solutions were ruled out in \cite{chae2007nonexistence} assuming 
sufficiently fast decay of the vorticity at infinity in physical space.
We see that the Dombre-Gilson scheme must be generalized in order to capture 
several scales. Possible generalizations are discussed in 
a separate paper~\cite{mailybaev2012b}.
 
\section{Summary}

Representation the blowup in the shell models by a traveling wave solution 
in the $\log k$ space suggested by 
Dombre and Gilson~\cite{dombre1998intermittency} is shown 
to be physically relevant for continuous hydrodynamic models. 
For compressible inviscid flows, a similar wave describes the 
universal scaling properties of shock formation. 

We showed that solutions of hyperbolic conservation laws describing 
compressible inviscid flows 
in one space dimension have universal scaling structure when approaching 
a finite-time singularity. The
limiting renormalized wave profile is described by a universal function in physical space,
or by a solitary wave moving with constant speed in logarithmic coordinates of
the Fourier space. This universal function is independent both of initial
conditions and of specific expressions for fluxes of conserved quantities.  
We also explained this phenomenon using the renormalization group theory.
Finally, we showed that the Dombre--Gilson scheme 
can be used for the study of blowup in incompressible inviscid flows.  

\section*{Acknowledgment} 
The author is grateful to C.M.~Dafermos, E.A.~Kuznetsov and D.~Marchesin 
for useful discussions and to the referee for suggesting using BKM criterion for inviscid shell models. This work was supported by CNPq under grant 477907/2011-3 and CAPES under grant PVE.

\section*{Appendix}

Using (\ref{eqS.3}), (\ref{eqS.6}) and (\ref{eqS.7}),  we find 
\begin{equation}
u_n(t) = -ik_n^{-1}v_n(t) = -ik_n^{-1}
\exp\left(\int_0^\tau A(\tau') d\tau'\right)W((n-s\tau)\log\lambda), 
\label{eqA.1}
\end{equation}
Note that expression (\ref{eqS.5b}) with $w_n$ from (\ref{eqS.7}) is invariant with respect to the shift $n \mapsto n+1$, $\tau \mapsto \tau+1/s$.
Hence, the function $A(\tau)$ is periodic with period
$1/s$, and we can write (\ref{eqA.1}) as
\begin{equation}
u_n(t) = -k_n^{-1}\lambda^{nz}F_1(\tau-n/s).
\label{eqA.3}
\end{equation}
Here the exponent $z$ with the function $F_1(\tau)$ are defined as 
\begin{equation}
z = \frac{1}{\log\lambda}\int_0^{1/s}A(\tau) d\tau,
\label{eqA.2}
\end{equation}
\begin{equation}
F_2(t-t_c) = F_1(\tau) 
= i\exp\left(\int_0^{\tau} A(\tau') d\tau'\right)W(-s\tau\log\lambda),
\label{eqA.4b}
\end{equation}
where the function $F_2$ will be used later, and $t$ is related to $\tau$ by (\ref{eqS.6}).

Let us find the time $t = t_n$ corresponding to $\tau-n/s$. Using (\ref{eqS.6}), we obtain
\begin{equation}
\begin{array}{c}
\displaystyle
t_n-t_c 
= -\int_{\tau-n/s}^\infty \exp\left(-\int_0^{\tau'} A(\tau'') d\tau''
\right)d\tau'
= -\int_{\tau}^\infty \exp\left(-\int_0^{\tau'-n/s} A(\tau'') d\tau''
\right)d\tau'
\\[17pt]
\displaystyle 
= -\lambda^{nz}\int_{\tau}^\infty \exp\left(-\int_0^{\tau'} A(\tau'') d\tau''
\right)d\tau' = \lambda^{nz}(t-t_c),
\end{array}
\label{eqA.4}
\end{equation}
where we used the change of variables $\tau' \mapsto \tau'+n/s$ and periodicity
of $A(\tau'')$. 
According to (\ref{eqA.4b}) and (\ref{eqA.4}), we have $F_1(\tau-n/s) =
F_2(t_n-t_c) = F_2(\lambda^{nz}(t-t_c))$. 
Using this function in (\ref{eqA.3}) with $k_n = k_0\lambda^n$ and taking $u_c = k_0^{-1}$, we
derive the expression 
\begin{equation}
u_n(t) 
= -k_0^{-z}k_n^{z-1}F_2\left(k_0^{-z}k_n^z(t-t_c)\right)
= -u_c(t_c-t)^{1/z-1}K^{z-1}F_2\left(-K^z\right),
\label{eqA.5}
\end{equation}
where $K = u_c(t_c-t)^{1/z}k_n$.
This expression with $F_2$ from (\ref{eqA.4b}) and $z$ from (\ref{eqA.2}) reduce to (\ref{eqS.2}), (\ref{eqF1}) with $\tau$ related to $t= t_c-K^z$ by (\ref{eqS.6}).

In the continuous case, we substitute $k_n$ by $k = e^\xi$ (so that $\lambda =
e$). We have $A = \mathrm{const}$ in (\ref{eqB.6}) for 
the traveling wave solution (\ref{eqB.7}). In this case we obtain expressions analogous to
(\ref{eqA.3})--(\ref{eqA.5}) in the form
\begin{equation}
u(t,k) = i k^{-2}v(t,\xi) = k^{z-2}F_1(\tau-\xi/s) 
= k^{z-2}F_2\left((t-t_c)k^z\right)
\label{eqA.6}
\end{equation}
with 
\begin{equation}
z = A/s, \quad
F_2(t-t_c) = F_1(\tau) 
= ie^{A\tau}W(-s\tau).
\label{eqA.7}
\end{equation}
For nonzero constant $A$, the first expression in (\ref{eqS.6}) yields the following relations 
\begin{equation}
t = t_c-\frac{e^{-A\tau}}{A},\quad
\tau = -\frac{\log(A(t_c-t))}{A}.
\label{eqA.7b}
\end{equation}
We reduce expression (\ref{eqA.6}) to the form (\ref{eqB.5}) with $u_c = 1$ 
by introducing the function 
\begin{equation}
F(k) = k^{z-2}F_2(-k^z).
\label{eqA.8}
\end{equation}
Note that an arbitrary factor $u_c > 0$ in (\ref{eqB.5}) reflects the scaling symmetry
of the Fourier transformed Burgers equation (\ref{eqB.1}) with $\nu = f = 0$. Using (\ref{eqA.7b}) in (\ref{eqA.7}) yields
\begin{equation}
F_2(t-t_c) = \frac{i}{A(t_c-t)}W\left(
\frac{\log(A(t_c-t)}{z}\right).
\label{eqA.9}
\end{equation}
Expression (\ref{eqB.5b}) for $F(k)$ follows from (\ref{eqA.8}), (\ref{eqA.9}).

\bibliographystyle{phjcp}
\bibliography{refs}

\end{document}